\documentclass[12pt]{iopart}

\usepackage{cite}
\usepackage[normalem]{ulem} 
\usepackage{soul} 
\usepackage{graphicx}
\usepackage{dcolumn}
\usepackage{bm}
\usepackage{euscript,stmaryrd}
\usepackage{xcolor}
\usepackage[utf8]{inputenc}
\usepackage[T1]{fontenc}
\usepackage{mathptmx}
\usepackage{epsf,amsmath,amssymb,verbatim,color,multirow,pifont,mathrsfs,soul,amsthm, wasysym,bm}
\usepackage{graphicx, upgreek}
\usepackage[us,12hr]{datetime}
\usepackage{xcolor}
\usepackage{arydshln}
\usepackage{hyperref}
\usepackage{arydshln}
\usepackage{graphicx}
\usepackage{mathtools}

\theoremstyle{definition}

\newcommand\redout{\bgroup\markoverwith
{\textcolor{red}{\rule[0.5ex]{2pt}{0.8pt}}}\ULon}

\newmuskip\pFqmuskip

\newcommand*\pFq[6][8]{%
  \begingroup 
  \pFqmuskip=#1mu\relax
  \mathcode`\,=\string"8000
  \begingroup\lccode`\~=`\,
  \lowercase{\endgroup\let~}\pFqcomma
  {}_{#2}F_{#3}{\left[\genfrac..{0pt}{}{#4}{#5};#6\right]}%
  \endgroup
}
\newcommand{\pFqcomma}{\mskip\pFqmuskip}

\usepackage{fancyhdr}
\usepackage{hyperref}
\hypersetup{
    colorlinks,
    citecolor=blue,
    filecolor=blue,
    linkcolor=blue,
    urlcolor=black
}
\usepackage{pdfrender}
\newcommand*{\boldgreek}[1]{%
  \textpdfrender{%
    TextRenderingMode=FillStroke,%
    LineWidth=.35pt,%
  }{#1}%
}

\begin{document}

\title[Volcano Transition in a System of Generalized Kuramoto Oscillators with Random Frustrated Interactions]{Volcano Transition in a System of Generalized Kuramoto Oscillators with Random Frustrated Interactions}

\author[1]{Seungjae Lee\footnote{Author to whom any correspondence should be addressed}}
\address{Physik-Department, Technische Universit\"at M\"unchen, James-Franck-Stra\ss e 1, 85748 Garching bei M\"unchen, Germany}

\ead{seungjae.lee@tum.de}

\author[2]{Yeonsu Jeong}
\address{Department of Applied Physics, Hanyang University ERICA, Ansan 15588, South Korea}

\author[3]{Seung-Woo Son}
\address{Department of Applied Physics, Hanyang University ERICA, Ansan 15588, South Korea}

\author[4]{Katharina Krischer}
\address{Physik-Department, Technische Universit\"at M\"unchen, James-Franck-Stra\ss e 1, 85748 Garching bei M\"unchen, Germany}

\vspace{10pt}
\begin{indented}
\item[]\today
\end{indented}

\begin{abstract} 
In a system of heterogeneous (Abelian) Kuramoto oscillators with random or `frustrated' interactions, transitions from states of incoherence to partial synchronization were observed. These so-called volcano transitions are characterized by a change in the shape of a local field distribution and were discussed in connection with an oscillator glass. In this paper, we consider a different class of oscillators, namely a system of (non-Abelian) SU(2)-Lohe oscillators that can also be defined on the 3-sphere, i.e., an oscillator is generalized to be defined as a unit vector in 4D Euclidean space. We demonstrate that such higher-dimensional Kuramoto models with reciprocal and nonreciprocal random interactions represented by a low-rank matrix exhibit a volcano transition as well. We determine the critical coupling strength at which a volcano-like transition occurs, employing an Ott-Antonsen ansatz. Numerical simulations provide additional validations of our analytical findings and reveal the differences in observable collective dynamics prior to and following the transition. Furthermore, we show that a system of unit 3-vector oscillators on the 2-sphere does not possess a volcano transition.

\end{abstract}

%
%
%
%
%

\section{\label{sec:intro}Introduction}

Numerous studies have explored emergent phenomena in complex systems across various disciplines~\cite{Strogatz_more}. An important type of system that has received significant attention during the last decades is coupled oscillators. Among them are large ensembles of attractively coupled oscillators that play a pivotal role in understanding transitions from incoherent states to partial synchronization~\cite{strogatz_sync,pikovksy_sync}. Given the similarity between the synchronization transition and ferromagnetic phase transitions~\cite{STROGATZ20001,kuramoto2003chemical}, Daido introduced a model involving Kuramoto phase oscillators with heterogenous natural frequencies and random frustrated interactions, featuring both attractive and repulsive couplings between oscillators~\cite{Daido_volcano,daido_volcano2}. It was conjectured that the resulting emergent phenomenon in this oscillator system might share characteristics with spin glasses~\cite{fischer1993spin,Castellani_2005}, motivating to coin this peculiar state an oscillator glass. Yet, based on further examinations, the definition of an oscillator glass has been a subject of ongoing debate and dispute~\cite{strogatz_volcano,pazo_volcano,glassy1,glassy2,glassy3,glassy4}. 

In 2018, Ottino-L\"offler and Strogatz put forward a mathematically tractable model, similar to Daido's approach~\cite{strogatz_volcano}. This model of heterogeneous Kuramoto phase oscillators was formulated with a symmetric, random interaction matrix with a tunable rank. Note that the interaction matrix of the model considered by Daido had a fixed, full rank. This system featured a so-called \textit{volcano transition}, which Daido suggested as indicative of an oscillator glass~\cite{Daido_volcano}. Above the volcano transition, an incoherent state loses its stability, and a so-called volcano phase, which is characterized by partial synchronization in a system with random frustrated interactions, becomes stable and thus observable. However, the volcano phase as a partially locked state with random frustrated interactions is not characterized by the usual Kuramoto order parameter, but rather requires measuring a local degree of synchrony, i.e., a local field defined below, for each oscillator. The volcano transition involves a change in the shape of the local field distribution, shifting from being concave downwards at the origin to concave upwards, i.e., a distribution of local fields that resembles a volcano shape. In contrast to Daido's conjecture, explorations of the model with low-ranked, random interaction matrices revealed that the partial synchrony beyond the volcano transition does not exhibit characteristics of an oscillator glass. For example, the volcano phase does not exhibit an algebraic relaxation of the Kuramoto order parameter, as expected for a glassy state, but rather displays an exponential relaxation~\cite{strogatz_volcano}. Thus, the relationship between an oscillator glass and a volcano phase continues to be a topic of ongoing discussion and requires further, thorough explorations. 

Realizing the significance of both nonreciprocal and random interactions for investigations related to real-world scenarios, Paz\'o and Gallego introduced a controllable symmetry/asymmetry parameter within the random interaction matrix to the model of Ottino-L\"offler and Strogatz ~\cite{pazo_volcano}. Thus, the random, frustrated interactions are rendered nonreciprocal by this additional parameter. Employing such random interactions, a volcano transition was observed and its relation to the symmetry-controlling parameter was discussed (see Sec.~\ref{sec:governing} for details). In both of the aforementioned models~\cite{strogatz_volcano,pazo_volcano}, the authors derived an analytical formula to identify the critical point at which a volcano transition takes place in the thermodynamic limit. This mathematical tractability relied on the so-called Ott-Antonsen (OA) ansatz, which can be invoked due to the sinusoidal coupling of the oscillators~\cite{OA1,OA2,WS_mobius,pikovsky_WS1,pikovsky_WS2}. Within the framework of the OA ansatz, the macroscopic dynamics of a system of heterogeneous Kuramoto phase oscillators can be described in terms of the so-called OA variables in the dimension-reduced manifold. Investigating the stability of incoherent states in terms of the OA variables, this approach enables the prediction of a critical point at which the volcano transition occurs.

Parallel to the studies mentioned above, the collective dynamics of ensembles of so-called generalized Kuramoto oscillators have attracted considerable attention~\cite{Tanaka_2014,g_KM1,g_KM2,g_KM3,g_KM4,g_KM5,g_KM6,g_KM7,short_chimera,Lee_2023}. In particular, Lohe detailed the dynamics of generalized Kuramoto models on a compact classical Lie group and provided insights into their notable characteristics~\cite{Lohe_2009,Lohe_WS}. Lohe's generalized Kuramoto oscillators can be understood as those defined on the $(D-1)$-sphere, with each oscillator represented as a unit vector in a $D$-dimensional Euclidean space. In contrast, standard Kuramoto phase oscillators are typically defined on the unit circle, with each oscillator represented as a unit vector embedded in the 2D plane (see Sec.~\ref{sec:governing} for details). Meanwhile, Barioni and Aguiar formulated the OA ansatz for higher-dimensional Kuramoto oscillators~\cite{barioni1,barioni2}. The authors proposed an expansion using higher-dimensional spherical harmonics for the oscillator density function in the thermodynamic limit, mirroring Ott and Antonsen's use of a Fourier series expansion for standard Kuramoto oscillators~\cite{OA1,OA2,WS_mobius,pikovsky_WS1,Omelchenko_2013,Omelchenko_2018,Laing_OA}. This ansatz yields the OA equation in a vector format, allowing for the description of the system's macroscopic dynamics in the dimension-reduced manifold.
 
In this paper, we study a higher-dimensional generalization of the aforementioned models. In Sec.~\ref{sec:governing}, we first revisit the model proposed by Paz\'o \textit{et al.}, thereby introducing the random, frustrated interactions. Then, we introduce a system of heterogeneous $\textrm{SU}(2)$-Lohe oscillators, coupled through reciprocal and nonreciprocal frustrated interactions. These oscillators are parameterized by unit 4-vectors, which allows us to investigate an ensemble of heterogeneous unit 4-vector oscillators on the 3-sphere and thus to identify a volcano transition. In Sec.~\ref{sec:theory}, we apply the higher-dimensional OA ansatz to this system, from which we obtain analytical predictions for the critical coupling strength as a function of the symmetry parameter. Section~\ref{subsec:4Dmodel} is dedicated to validating our analytical results through numerical simulations with large ensemble sizes. Furthermore, in Sec.~\ref{subsec:3Dmodel}, we discuss a system of 3D unit vector oscillators for comparison with our main result of the 4D model. Finally, we summarize our results and offer an outlook for future research in Sec.~\ref{sec:Conlcusion}.

\section{\label{sec:governing}Governing Equations with Random Frustrated Interactions}

For the exploration of a volcano transition, one can consider the system of heterogeneous Kuramoto phase oscillators introduced in Ref.~\cite{pazo_volcano}. Each oscillator $\phi_i \in \mathbb{T}:=[-\pi,\pi]$ for $i \in [N]:= \{1,2,...,N\}$ is governed by
\begin{flalign}
    \dot{\phi}_i &= \omega_i +J~\text{Im} \bigg[ P_i(t) e^{-\textrm{i}\phi_i} \bigg] \notag \\
    &=\omega_i + \frac{J}{N}\sum_{j=1}^{N}M_{ij}\sin(\phi_j-\phi_i), \label{eq:pazo_governing}
\end{flalign} where $N$ is the number of oscillators, $J \in \mathbb{R}_+$ is the (global) coupling strength, and $P_i(t)$ denotes complex local forcing fields (see below). The natural frequency ($\omega_i \in \mathbb{R}$) of oscillator $i$ is selected from any unimodal distribution, which we take to be centered at zero, exploiting the rotational symmetry of Eq.~(\ref{eq:pazo_governing}). More precisely, we employ a Gaussian distribution
\begin{equation}
 g(\omega) = \frac{1}{\Delta \sqrt{2\pi}}e^{-\frac{\omega^2}{2\Delta^2}} ,  \label{eq:gaussian_dist}
\end{equation} where $\Delta$ is the standard deviation, which is fixed at $\Delta=1$ throughout this paper. Paz\'o \textit{et al.} observed a volcano transition with an interaction matrix $(M_{ij}) = \bold{M} \in \mathbb{R}^{N \times N}$ which characterizes quenched random disorder in terms of interactions between oscillators. The interaction matrix was obtained as follows~\cite{pazo_volcano}: Each oscillator is assigned two time-independent random vectors $\bold{u}_i, \bold{v}_i \in \{\pm1\}^L$ for $i \in [N]$, where $L \ge 2 $ is the dimension of the interaction vectors. Their components are selected randomly and independently as $\pm 1$ with probability $0.5$, and the matrix of random frustrated interactions is defined as
\begin{equation}
    \bold{M} := \frac{1}{2} \big[ (1+\eta) \bold{S} +(1-\eta)\bold{A} \big], \label{eq:random_matrix}
\end{equation} where $\eta \in [-1,1]$ is a bifurcation parameter controlling the weight between $\bold{S}$ and $\bold{A}$ in the interaction matrix $\bold{M}$. Here, the symmetric matrix $\bold{S}$ is defined by inner products $S_{ij} := \langle \bold{u}_i, \bold{u}_j \rangle - \langle \bold{v}_i, \bold{v}_j \rangle$ that characterizes a reciprocal interaction whereas the asymmetric matrix $\bold{A}$ given by $A_{ij} :=  \langle \bold{u}_i, \bold{v}_j \rangle - \langle \bold{v}_i, \bold{u}_j \rangle$. It determines an anti-reciprocal interaction between oscillators $i$ and $j$. Both satisfy $S_{ii}=A_{ii}=0$ for $i=1,...,N$ (no self-interaction). It is worth noting that in Ref.~\cite{strogatz_volcano} the authors only investigated symmetric, reciprocal interactions between oscillators, corresponding to $\eta=1$ in Eq.~(\ref{eq:random_matrix}). However, for $\eta<1$, the interaction is rendered nonreciprocal. Additionally, the interaction matrix encompasses both attractive ($M_{ij}>0$) and repulsive ($M_{ij}<0$) couplings. All the detailed properties of the above interaction matrix are summarized in Refs.~\cite{pazo_volcano,strogatz_volcano}. Throughout this paper, we maintain a constant value of $L=2$. This choice adheres to the condition $L \ll \log_2N$. This also ensures that, for the sake of analytical tractability, the interaction matrix $(M_{ij})$ is intentionally configured as a low-ranked quenched disorder, rather than being fully ranked, i.e., $\text{rank}( \bold{M})=2L$ except for $\eta=0$~\cite{pazo_volcano}. 

In the above setup, the detection of a volcano transition can be accomplished by assessing radial distributions of complex local fields, which are defined by
\begin{equation}
    P_i = r_i e^{\textrm{i}\psi_i} = \frac{1}{N}\sum_{j=1}^N M_{ij} e^{\textrm{i}\phi_j}. \label{eq:local_fields_pazo}
\end{equation} 
Note that the global Kuramoto order parameter is defined by
\begin{equation}
    \Gamma(t):=\frac{1}{N}\sum_{j=1}^N e^{\textrm{i}\phi_j(t)} \in \mathbb{C}, \label{eq:global_order_pazo}
\end{equation}
which cannot be used to observe a volcano transition since it remains close to zero for all coupling strength $J$. This explains a prominent difference between well-known partial synchronization in a system of globally coupled oscillators and a volcano phase of oscillators with random frustrated interactions. The latter can be detected by measuring a local degree of coherence on the connectivity topology whereas the former is described by the global order parameter. At the transition, the radial distribution $h(r_i)$ of the local fields becomes concave up at the origin (the most probable value of $r_i$ is greater than zero) from being concave down (the most probable value of $r_i$ is zero)~\cite{pazo_volcano,strogatz_volcano,Daido_volcano}. Furthermore, employing the OA ansatz~\cite{OA1,OA2,WS_mobius,pikovsky_WS1,pikovsky_WS2}, an analytical prediction of the critical coupling strength $J_c$ was derived in Refs.~\cite{pazo_volcano,strogatz_volcano}, using the linear stability analysis of incoherent states in the thermodynamic limit:
\begin{equation}
    J_c = \frac{1}{\sqrt{\eta}} \times \frac{2}{\pi g(0)}, \label{eq:critical_pazo}
\end{equation} which is a function of the parameter $\eta$. Above $J_c$, the incoherent state is destabilized. Hence, it was shown that a volcano transition can occur in a system of heterogeneous Kuramoto oscillators with nonreciprocal as well as reciprocal interactions~\cite{pazo_volcano}. Moreover, Eq.~(\ref{eq:critical_pazo}) demonstrates that the likelihood of observing the volcano transition decreases as the symmetry parameter $\eta$ becomes smaller, and eventually goes to zero as $\eta \rightarrow 0^+$.

As outlined above, the previous explorations of the volcano transition primarily focused on systems of heterogenous Kuramoto phase oscillators, employing Eq.~(\ref{eq:pazo_governing}) or similar systems~\cite{pazo_volcano,strogatz_volcano,Daido_volcano}. In this paper, we consider a system of so-called generalized Kuramoto oscillators proposed by Lohe (for details of the below description, see Ref.~\cite{Lohe_2009}) that consists of non-Abelian Kuramoto oscillators where each oscillator $U_i \in \textrm{SU}(2)$ is governed by
\begin{flalign}
    \textrm{i} \dot{U}_i {U_{i}}^\dagger = H_i - \frac{\textrm{i}J}{2N}\sum_{j=1}^{N}M_{ij} \big( U_i {U_j}^\dagger - U_j {U_i}^\dagger \big). \label{eq:Lohe_governing}
\end{flalign} Here, $\dagger$ indicates the Hermitian conjugate. Note that in a broader context, one has the flexibility to consider various classical compact Lie groups for this generalization, e.g., $U_i \in \textrm{U}(n)$ or $\textrm{SO}(n)$~\cite{Lohe_2009}. However, our focus in this paper is exclusively on the subgroup $\textrm{SU}(2)$. In Eq.~(\ref{eq:Lohe_governing}), one can choose the local dynamics to be dictated by a time-independent Hermitian matrix with zero trace, denoted as $H_i \in \mathfrak{su}_2 $ (Lie algebra). Then, the right-hand side of Eq.~(\ref{eq:Lohe_governing}) retains Hermitian properties. This implies that the dynamics of the oscillators are restricted to evolve on the compact manifold of $\textrm{SU}(2)$ as long as the initial condition $U_i(0) \in \textrm{SU}(2)$ holds. As a consequence, $U_i{ U_i}^\dagger$ is a constant of motion and $U_i(t) \in \textrm{SU}(2)$ for all $t>0$. In essence, the local dynamics is nothing but a finite-dimensional Schr\"odinger's equation, as it can be expressed by $\textrm{i} \dot{U}_i = H_i U_i$. It is noteworthy that the Abelian Kuramoto model in Eq.~(\ref{eq:pazo_governing}) can be revisited using $U_i = e^{-\textrm{i}\phi_i} \in \textrm{U}(1)$ and setting $H_i = \omega_i$ for $i \in [N]$. Furthermore, as pointed out in Ref.~\cite{Lohe_2009}, the equation exhibits a chiral $\textrm{SU}(2)\times \textrm{SU}(2)$ covariance. This can be seen when considering the transformations $U_i \mapsto S U_i T$ and $H_i \mapsto S H_i S^\dagger$, with $S,T$ denoting constant unitary matrices. In such cases, $S U_i T$ also satisifies Eq.~(\ref{eq:Lohe_governing}) if $U_i$ is a solution. 

Given our exclusive focus on the $\textrm{SU}(2)$ model, we can parameterize $U_i \in \textrm{SU}(2)$ using a real 4-vector with unit magnitude $\bold{x}_i = (x_i^1,x_i^2,x_i^3,x_i^4)^\intercal \in S^3 := \{ \bold{x} \in \mathbb{R}^4 | \langle \bold{x},\bold{x} \rangle=1 \} $ on the 3-sphere since the group $\textrm{SU}(2)$ is diffeomorphic to the 3-sphere~\cite{gilmore2006lie}. This parameterization reads for $i \in [N]$
\begin{flalign}
    U_i = \textrm{i} \sum_{k=1}^3 x^k_i \sigma_k + x^4_i I_2 = \begin{pmatrix}
x^4_i+\textrm{i} x^3_i & x^2_i + \textrm{i} x^1_i  \\
-x^2_i + \textrm{i} x^1_i & x^4_i - \textrm{i} x^3_i 
\end{pmatrix}, 
\label{eq:parameterization_unitary}
\end{flalign} where $\sigma_k$ for $k=1,2,3$ are Pauli matrices and $I_n \in \mathbb{R}^{n \times n}$ is the identity matrix~\cite{Lohe_2009}. Substituting Eq.~(\ref{eq:parameterization_unitary}) into Eq.~(\ref{eq:Lohe_governing}), one can obtain the governing equations of the unit 4-vectors on $S^3$ that read
\begin{flalign}
    \dot{\bold{x}}_i = \Omega_i \bold{x}_i + \frac{J}{N}\sum_{j=1}^{N}M_{ij}\big( \bold{x}_j - \langle \bold{x}_i, \bold{x}_j \rangle \bold{x}_i \big).  
    \label{eq:vector_governing}
\end{flalign} 
Also, Eq.~(\ref{eq:vector_governing}) guarantees that if $\| \bold{x}_i(0) \|=1$ holds, then $\| \bold{x}_i(t) \|=1$ for $t>0$, namely, $\| \bold{x}\| := \sqrt{\langle\bold{x},\bold{x}\rangle}$ is a constant of motion. Moreover, if we restrict the system to 2-vector oscillators ($\bold{x}_i = (\cos\phi_i,\sin\phi_i)^\intercal \in S^{1}$), then we regain the usual (2D) Kuramoto model as in Eq.~(\ref{eq:pazo_governing}). In Eq.~(\ref{eq:vector_governing}), the local dynamics is governed by a real, anti-symmetric natural frequency matrix for each oscillator, i.e., $\Omega_i^\intercal = - \Omega_i$ for $i \in [N]$ where $\intercal$ denotes the transpose of a matrix. These natural frequency matrices, as described in Eq.~(19) of Ref.~\cite{Lohe_2009}, do not represent the most general form of elements of $\mathfrak{so}_4$. Instead, they span $\mathfrak{su}_2$, corresponding to an $\textrm{SU}(2)$ subgroup of the $\textrm{SO}(4)$ group. This correspondence arises from the local isomorphism $\textrm{SO}(4) \cong \textrm{SU}(2) \times \textrm{SU}(2)$ and the presence of the chiral covariance~\cite{Lohe_2009}. Nevertheless, for the sake of generality, we will utilize the most general form of elements from $\mathfrak{so}_4$ to assign each oscillator an anti-symmetric natural frequency matrix. Thus, a natural frequency matrix of each oscillator for our system is given by
\begin{equation}
   \Omega_i = \begin{pmatrix}
0 &  -\omega_i^6 & \omega_i^5  & -\omega_i^4 \\
\omega_i^6 & 0 & -\omega_i^3 & \omega_i^2 \\
-\omega_i^5 & \omega_i^3 & 0  & -\omega_i^1 \\
\omega_i^4 & -\omega_i^2 & \omega_i^1 & 0
\end{pmatrix} \label{eq:nat_freq_matrix_4D}
\end{equation} for $i\in [N]$ and satisfies $\Omega_i^\intercal = -\Omega_i$. We also assume that the natural frequency matrices are distributed according to a distribution $G(\bm{\Omega})$. As in Ref.~\cite{PRX_generalized}, we here choose the natural frequency matrix distribution $G(\bm{\Omega})$ as follows: each component $ \omega_i^a $ for each $a=1,...,6$ is selected randomly and independently from a Gaussian distribution in Eq.~(\ref{eq:gaussian_dist}) with $\Delta=1$. This method allows us to exploit rotational invariance and so the anti-symmetric matrices are set to have zero mean. This fact will be useful later in Sec.~\ref{subsec:critical_point}. Lastly, the random frustrated interactions $M_{ij}$ in Eq.~(\ref{eq:vector_governing}) are defined in Eq.~(\ref{eq:random_matrix}).

Also in this system, the global Kuramoto order parameter, which is here defined as a center of mass of oscillators on $S^3$, namely,
\begin{equation}
    \bm{\Gamma}(t) := \frac{1}{N}\sum_{j=1}^{N} \bold{x}_j(t) \label{eq:global_order}
\end{equation} remains close to a zero vector for $J\ge 0$ and so it is not indicative of a volcano transition. Therefore, we define a local field vector as
\begin{equation}
    \bold{P}_i = (P_i^1, P_i^2, P_i^3, P_i^4 )^\intercal := \frac{1}{N}\sum_{j=1}^N M_{ij} \bold{x}_j \in \mathbb{R}^4 \label{eq:local_field_vector}
\end{equation} for $i \in [N]$, similarly to Eq.~(\ref{eq:local_fields_pazo}). Given its four-dimensional distribution, we will measure its partial information by examining distributions on individual planes in $\mathbb{R}^4$ such as $h(x^1_i, x^2_i)$ or $h(x_i^3, x_i^4)$. This approach provides an effective means to observe a volcano transition in our system. See numerical results in Sec.~\ref{subsec:4Dmodel} for details.

\section{\label{sec:theory}Theoretical Result: Critical Coupling Strength}

In this section, we derive an analytical expression for the critical coupling strength $J_c$ in the non-Abelian Kuramoto model (\ref{eq:vector_governing}) at which the volcano transition occurs, i.e., at which the incoherent state loses its stability.

\subsection{\label{subsec:higher_OA}Higher-dimensional Ott-Antonsen Ansatz}

We start by applying the so-called higher-dimensional OA ansatz, recently introduced in Refs.~\cite{barioni1,barioni2}, to our system (\ref{eq:vector_governing}) with random frustrated interactions. Although we are interested in 4D oscillators, we here consider any even-dimensional oscillators defined on the $(D-1)$-sphere for $D \ge 2$, i.e., $\bold{x}_i \in S^{D-1}$, in order to provide a general description. The odd-dimensional cases are considered in Sec.~\ref{subsec:3Dmodel}.

The governing equations (\ref{eq:vector_governing}) of the non-Abelian Kuramoto oscillators describe a finite-sized ensemble. However, to apply the OA ansatz, we need to consider the thermodynamic limit $N \rightarrow \infty$. The state function of this system is the density of oscillators $f(\hat{\bold{r}}, \Omega, \bold{w};t)$, specified by the continuous natural frequencies  $\Omega$ and the $2^{2L}$ interaction vectors $\bold{w} := (\bold{u}^\intercal,\bold{v}^\intercal)^\intercal \in \mathbb{R}^{2L}$ which are discrete parameters for a continuous unit vector oscillator $\hat{\bold{r}} = \hat{\bold{r}}(\theta_1, \cdots, \theta_D) \in S^{D-1}$. Since the oscillator density function is defined on the sphere $S^{D-1}$, it can be expanded by higher-dimensional spherical harmonics~\cite{barioni2} such that
\begin{flalign}
    f(\hat{\bold{r}}, \Omega, \bold{w};t) = G(\bm{\Omega}) \sum_{\ell_{D-1}=0}^{\infty} \sum_{\ell_{D-2}=0}^{\ell_{D-1}} \cdots \sum_{m=-\ell_2}^{\ell_2} f_{\ell_2 \cdots \ell_{D-1}}^{m}(\bm{\Omega},\bold{w};t) Y_{\ell_2 \cdots \ell_{D-1}}^{m}(\hat{\bold{r}}) \label{eq:spherical_expansion}
\end{flalign} similar as the oscillator density function for the Abelian Kuramoto oscillators can be expanded in Fourier series~\cite{OA1,WS_mobius,pikovsky_WS1,OA2,pikovsky_WS2}. Furthermore, in the thermodynamic limit, the continuous local field vector is defined by
\begin{flalign}
    \bold{P}(\bold{w},t) := \frac{1}{2^{2L}}\sum_{\bold{w}'}\tilde{\mathrm{M}}(\bold{w},\bold{w}')\int \int_{S^{D-1}} \hat{\bold{r}} f(\hat{\bold{r}}, \Omega, \bold{w}';t)d\bm{\Omega}d^DS ~, \label{eq:local_field_continuous}
\end{flalign} where $\tilde{\mathrm{M}}(\bold{w},\bold{w}') $ is the $(\bold{w},\bold{w}')$-th matrix element of  $ \tilde{\textbf{M}} \in \mathbb{R}^{2^{2L} \times 2^{2L}}$~\cite{pazo_volcano}. Due to the conservation of the number of oscillators, the oscillator density function is governed by the continuity equation that reads~\cite{barioni1,barioni2}
\begin{flalign}
    \frac{\partial}{\partial t} f(\hat{\bold{r}}, \Omega, \bold{w};t) + \langle \nabla_S f -(D-1) f \hat{\bold{r}}  , J \bold{P} \rangle + \langle \bm{\Omega} \hat{\bold{r}}, \nabla_S f \rangle =0 ~, \label{eq:continuity_equation}
\end{flalign} where $\nabla_S f$ denotes the component of the gradient of $f$ along the surface $S^{D-1}$ of the unit ball~\cite{PRX_generalized}.

The OA ansatz for the usual 2D Kuramoto model assumes that the $n$-th Fourier coefficient is given by the $n$-th power of the first coefficient. This first coefficient constitutes the OA variable and describes the system's macroscopic dynamics~\cite{OA1,WS_mobius,pikovsky_WS1}. Likewise, Barioni \textit{et al.} suggested \textit{the higher-dimensional Ott-Antonsen ansatz} constructed similarly to the 2D case. Following their ansatz, we assume a manifold where the coefficients in Eq.~(\ref{eq:spherical_expansion}) are given by
\begin{equation}
    f_{\ell_2 \cdots \ell_{D-1}}^{m}(\bm{\Omega},\bold{w};t) = \rho^{\ell_{D-1}} \overline{Y^{m}_{\ell_2 \cdots \ell_{D-1}}}(\hat{\boldgreek{\bm{\rho}}})~, \label{eq:OA_ansatz}
\end{equation} where the vector $\boldgreek{\bm{\rho}} = \rho \hat{\boldgreek{\bm{\rho}}}(\Theta_1,\cdots,\Theta_D) \in \mathbb{R}^D$ is an OA vector with $\rho = \rho(\bm{\Omega},\bold{w},t) \in [0,1)$ and $\hat{\boldgreek{\bm{\rho}}} \in S^{D-1}$~\cite{barioni2}. The symbol with an overbar indicates the complex conjugate. Plugging Eq.~(\ref{eq:OA_ansatz}) into Eq.~(\ref{eq:spherical_expansion}), the oscillator density is written as a higher-dimensional normalized Poisson kernel such that
\begin{flalign}
   f(\hat{\bold{r}}, \Omega, \bold{w};t) &= G(\bm{\Omega}) \sum_{\ell_{D-1}=0}^{\infty} \sum_{\ell_{D-2}=0}^{\ell_{D-1}} \cdots \sum_{m=-\ell_2}^{\ell_2} \rho^{\ell_{D-1}} \overline{Y^{m}_{\ell_2 \cdots \ell_{D-1}}}(\hat{\boldgreek{\bm{\rho}}})  Y_{\ell_2 \cdots \ell_{D-1}}^{m}(\hat{\bold{r}}) \notag \\
   &=G(\bm{\Omega}) \frac{\Gamma(D/2)}{2\pi^{D/2}} \frac{1-\rho^2}{(1+\rho^2-2\rho\langle \hat{\bold{r}}, \hat{\boldgreek{\bm{\rho}}}\rangle)^{D/2}} \notag \\
   &=G(\bm{\Omega}) \frac{1}{S_D}\frac{1-\| \boldgreek{\bm{\rho}} \|^2}{\| \boldgreek{\bm{\rho}} - \hat{\bold{r}} \|^D} ~, 
   \label{eq:OA_manifold}
\end{flalign} where $S_D:=\frac{2\pi^{D/2}}{\Gamma(D/2)}$ as pointed out in Refs.~\cite{Lee_2023,barioni2}. Hence, in the higher-dimensional OA manifold, the local field vector defined in Eq.~(\ref{eq:local_field_continuous}) reads
\begin{flalign}
    \bold{P}(\bold{w},t) = \frac{1}{2^{2L}}\sum_{\bold{w}'} \tilde{\mathrm{M}}(\bold{w},\bold{w}')\int G(\bm{\Omega}) \boldgreek{\bm{\bm{\rho}}}(\bm{\Omega},\bold{w}',t) d\bm{\Omega} \label{eq:local_field_OA}
\end{flalign} for each interaction vector $\bold{w}$ since the normalized Poisson kernel leads to
\begin{flalign}
    \int_S \hat{\bold{r}} f(\hat{\bold{r}},\bm{\Omega},\bold{w}';t)d^DS &= \frac{1}{S_D} G(\bm{\Omega}) \int_S \hat{\bold{r}} ~  \frac{1-\| \boldgreek{\bm{\rho}} \|^2}{\| \boldgreek{\bm{\rho}} - \hat{\bold{r}} \|^D} d^DS \notag \\
    &= G(\bm{\Omega}) \frac{1-\| \boldgreek{\bm{\rho}} \|^2}{S_D} \int_{-1}^{1}\int_{\|\bold{n}\|=\sqrt{1-\eta^2}}\frac{\eta\hat{\boldgreek{\bm{\rho}}} +\bold{n} }{(1 + \|\boldgreek{\bm{\rho}}\|^2-2 \langle \boldgreek{\bm{\rho}}, \hat{\bold{r}} \rangle)^{D/2}} d\bold{n} \frac{d\eta}{\sqrt{1-\eta^2}} \notag \\
    &= G(\bm{\Omega}) \frac{S_{D-1}}{S_D} (1-\| \boldgreek{\bm{\rho}} \|^2) \hat{\boldgreek{\bm{\rho}}} \int_{-1}^{1} \frac{\eta}{(1 + \|\boldgreek{\bm{\rho}}\|^2-2 \|\boldgreek{\bm{\rho}}\| \eta )^{D/2}} \frac{(1-\eta^2)^{\frac{D-2}{2}}}{\sqrt{1-\eta^2}} d \eta \notag \\ 
    &= G(\bm{\Omega}) \boldgreek{\bm{\rho}}(\bm{\Omega},\bold{w}',t) ~,
\end{flalign} where $\hat{\boldgreek{\bm{\rho}}} = \frac{\boldgreek{\bm{\rho}}}{\|\boldgreek{\bm{\rho}}\|}$ is a normalized OA vector and $\hat{\bold{r}} = \eta \hat{\boldgreek{\bm{\rho}}} + \bold{n} $ with $\langle \boldgreek{\bm{\rho}}, \bold{n} \rangle = 0 $~\cite{Lee_2023,Tanaka_2014}. Finally, substituting Eq.~(\ref{eq:local_field_OA}) and Eq.~(\ref{eq:OA_manifold}) into the continuity equation (\ref{eq:continuity_equation}), one can derive the higher-dimensional OA equation~\cite{barioni2}:
\begin{flalign}
    \frac{d}{dt}\boldgreek{\bm{\rho}}(\bm{\Omega},\bold{w},t) = \bm{\Omega} \boldgreek{\bm{\rho}} + \frac{J}{2}\big( 1+ \| \boldgreek{\bm{\rho}} \|^2 \big)  \bold{P}(\bold{w},t) - J \langle \boldgreek{\bm{\rho}} , \bold{P} \rangle \boldgreek{\bm{\rho}}. \label{eq:final_OA_equation}
\end{flalign} Note that Lohe obtained a similar dimension-reduced governing equation for the identical higher-dimensional oscillators using the higher-dimensional Watanabe-Strogatz transformation~\cite{Lohe_WS}.

\subsection{\label{subsec:critical_point}The Critical Coupling Strength}

In this subsection, we obtain the critical coupling strength exploiting the higher-dimensional OA ansatz in Sec.~\ref{subsec:higher_OA}. The volcano transition occurs at the parameter value at which the incoherent state loses its stability. Therefore, we determine the linear stability of the incoherent state, as also done in Refs.~\cite{pazo_volcano,strogatz_volcano}. Let us consider an infinitesimal perturbation vector $\bm{\beta}$ with $\| \bm{\beta} \| \ll 1$ off the incoherent state $\boldgreek{\bm{\rho}}_0 = \bold{0}$. Substituting $\boldgreek{\bm{\rho}} (\bm{\Omega},\bold{w},t) = \boldgreek{\bm{\rho}}_0 + \bm{\beta}(\bm{\Omega},\bold{w})e^{\lambda t}$ into Eq.~(\ref{eq:final_OA_equation}), we obtain the following linearized equation:
\begin{flalign}
    \lambda \bm{\beta} (\bm{\Omega},\bold{w}) = \bm{\Omega} \bm{\beta} + \frac{J}{2^{2L+1}}\sum_{\bold{w}'} \tilde{\mathrm{M}}(\bold{w},\bold{w}') \int G(\bm{\Omega}')\bm{\beta}(\bm{\Omega}',\bold{w}') d\bm{\Omega}' \label{eq:perturbation}
\end{flalign} for each interaction vector $\bold{w}$. As described in Refs.~\cite{PRX_generalized,barioni2}, there exists a real orthogonal matrix $\bold{O}$ that transforms the anti-symmetric natural frequency matrix $\bm{\Omega}$ into a block-diagonalized form~\cite{golub1983matrix}. Under the transformation $\bm{\Omega} \mapsto \bold{O}^\intercal \bm{\Omega} \bold{O}$, the natural frequency matrix reads
\begin{equation}
   \bold{D} := \bold{O}^\intercal \bm{\Omega} \bold{O} = \begin{pmatrix}
\bold{W}_1 &  &  \\
 & \ddots & \\
 &  &  \bold{W}_{D/2}
\end{pmatrix} ~,
\end{equation} where all the off-diagonal blocks are zero matrices and each diagonal block is given by
    $\bold{W}_i = \begin{pmatrix}
0 & \omega_i \\
 -\omega_i & 0
\end{pmatrix}$
for $i=1,...,D/2$. Applying the transformation $\bm{\beta} \mapsto \bold{O} \bm{\beta}$ in Eq.~(\ref{eq:perturbation}), we obtain
\begin{flalign}
    \lambda \bm{\beta}(\bm{\Omega},\bold{w}) &= \bold{O}^\intercal \bm{\Omega} \bold{O} \bm{\beta} +\frac{J}{2^{2L+1}} \sum_{\bold{w}'} \tilde{\mathrm{M}}(\bold{w},\bold{w}') \int G(\bm{\Omega}')\bm{\beta}(\bm{\Omega}',\bold{w}') d\bm{\Omega}' \notag \\
    &= \bold{D} \bm{\beta} +\frac{J}{2^{2L+1}} \sum_{\bold{w}'} \tilde{\mathrm{M}}(\bold{w},\bold{w}') \int G(\bm{\Omega}')\bm{\beta}(\bm{\Omega}',\bold{w}') d\bm{\Omega}'  \label{eq:derivation1} \\
&= \bold{D} \bm{\beta}  +\frac{J}{2^{2L+1}} \sum_{\bold{w}'} \tilde{\mathrm{M}}(\bold{w},\bold{w}') \int \cdots \int g(\{\omega_i\})\bm{\beta}(\{\omega_i\},\bold{w}') F(\bold{O}')d\omega_1 \cdots d\omega_{D/2} d\bold{O}'. \notag
\end{flalign} 
In Eq.~(\ref{eq:derivation1}), one can choose $G(\bm{\Omega}) = g(\{\omega_i\}) F(\bold{O})$ due to the rotational invariance of $G(\bm{\Omega})$. Here, $g(\{\omega_i\})=g(\omega_1,\cdots,\omega_{D/2})$ is a joint distribution of the frequencies $\{ \omega_i \}$ and $F(\bold{O})$ denotes a uniform distribution of orthogonal matrices (see Appendix C of Ref.~\cite{PRX_generalized}). Considering each subspace with $\bold{W}_i$ and using a complex representation, we rewrite the perturbation vector as
\begin{equation}
    \bm{\beta}(\{\omega_i\},\bold{w})=\begin{pmatrix}
b_1(\{\omega_i\},\bold{w}) \\
 \vdots \\
b_{D/2}(\{\omega_i\},\bold{w})
\end{pmatrix}~,
\end{equation} where $b_i \in \mathbb{C}^1$ for $i=1,...,D/2$. Then, the linearized equation (\ref{eq:derivation1}) reads for each $\bold{w}$
\begin{flalign}
    \lambda \begin{pmatrix}
b_1(\{\omega_i\},\bold{w}) \\
 \vdots \\
b_{D/2}(\{\omega_i\},\bold{w})
\end{pmatrix} &= \begin{pmatrix}
-\textrm{i}\omega_1 &  &  \\
 & \ddots &  \\
 &  &  -\textrm{i}\omega_{D/2}
\end{pmatrix} \begin{pmatrix}
b_1 \\
 \vdots \\
b_{D/2}
\end{pmatrix} \label{eq:derivation2} \\
&+\frac{J}{2^{2L+1}} \sum_{\bold{w}'}\tilde{\mathrm{M}}(\bold{w},\bold{w}') \int_{\mathbb{R}} \cdots \int_{\mathbb{R}} g(\{\omega_i\})\begin{pmatrix}
b_1(\{\omega_i\},\bold{w}') \\
 \vdots \\
b_{D/2}(\{\omega_i\},\bold{w}')
\end{pmatrix} d\omega_1 \cdots d\omega_{D/2}. \notag
\end{flalign} 
We define now
$    \hat{b}_n(\bold{w}) = \int_{\mathbb{R}} \cdots \int_{\mathbb{R}} g(\omega_1, \cdots, \omega_{D/2}) b_n(\omega_1, \cdots, \omega_{D/2},\bold{w})d\omega_1 \cdots d\omega_{D/2} $
for $n=1,...,D/2$. This allows us to algebraically rearrange Eq.~(\ref{eq:derivation2}) and integrate it over the joint distribution of the frequencies on both sides. We then arrive at
\begin{flalign}
    \begin{pmatrix}
\hat{b}_1(\bold{w}) \\
 \vdots \\
\hat{b}_{D/2}(\bold{w})
\end{pmatrix} &= \frac{J}{2^{2L+1}} \int_\mathbb{R} \cdots \int_\mathbb{R} g(\omega_1, \cdots, \omega_{D/2}) \begin{pmatrix}
\frac{1}{\lambda+\textrm{i}\omega_1} &  &  \\
 & \ddots &  \\
 &  & \frac{1}{\lambda+\textrm{i}\omega_{D/2}}
\end{pmatrix}d\omega_1 \cdots d\omega_{D/2} \notag \\
& \times\sum_{\bold{w}'}\tilde{\mathrm{M}}(\bold{w},\bold{w}') \begin{pmatrix}
\hat{b}_1(\bold{w}') \\
 \vdots \\
\hat{b}_{D/2}(\bold{w}')
\end{pmatrix}. \label{eq:derivation3}
\end{flalign} Note that at the critical point, the real part of the eigenvalue $\lambda$ goes to zero~\cite{pazo_volcano}. To exploit this, let us consider the first component $\hat{b}_1(\bold{w})$. We obtain
\begin{flalign}
    \hat{b}_1(\bold{w}) &= \frac{J}{2^{2L+1}} \sum_{\bold{w}'}\tilde{\mathrm{M}}(\bold{w},\bold{w}')\hat{b}_1(\bold{w}')  \int_\mathbb{R} d\omega_2 \cdots \int_\mathbb{R} d\omega_{D/2} \int d\omega_1 \frac{g(\omega_1,\cdots,\omega_{D/2})}{\lambda + \textrm{i}\omega_1}  \notag \\
    &= \frac{J}{2^{2L+1}} \sum_{\bold{w}'}\tilde{\mathrm{M}}(\bold{w},\bold{w}')\hat{b}_1(\bold{w}') \int_\mathbb{R} d\omega_2 \cdots \int_\mathbb{R} d\omega_{D/2}  \pi g(0,\omega_2, \cdots,\omega_{D/2})   \notag \\
    &=\pi \tilde{g}(0)\frac{J}{2^{2L+1}} \sum_{\bold{w}'}\tilde{\mathrm{M}}(\bold{w},\bold{w}')\hat{b}_1(\bold{w}') ~, 
\end{flalign} where the value of $\tilde{g}(0)$ is given by~\cite{PRX_generalized}
\begin{equation}
    \tilde{g}(0) := \int_\mathbb{R} d\omega_2 \cdots \int_\mathbb{R} d\omega_{D/2} ~ g(0,\omega_2,\cdots,\omega_{D/2}) = \frac{1}{D} \sqrt{\frac{2}{\pi}}\sum_{n=0}^{D/2-1}\frac{(2n)!}{2^{2n}(n!)^2}
\end{equation} for $D \ge 2$ if we consider the zero mean distribution of natural frequency matrices. Since all the other components obey the same equation (\ref{eq:derivation3}), we obtain a matrix form of an eigenvalue equation that determines the stability of the incoherent state:
\begin{equation}
    \bigg( \frac{J_c}{2^{2L+1}} \pi \tilde{g}(0)  \tilde{\textbf{M}} - \bm{I} \bigg) \hat{\bold{b}} = \bold{0} ~,
    \label{eq:derivation4} 
\end{equation} where $\bm{I}$ is the identity matrix. The nontrivial solutions of Eq.~(\ref{eq:derivation4}) ($\hat{\bold{b}} \neq \bold{0}$) are given by eigenvectors of $\tilde{\textbf{M}}$ corresponding to nonzero eigenvalues~\cite{pazo_volcano}. The nonzero eigenvalues of $\tilde{\textbf{M}}$ are given by $\sqrt{\eta}2^{2L}$ and $-\sqrt{\eta}2^{2L}$ with multiplicity $L$~\cite{pazo_volcano,strogatz_volcano}. Thus, the volcano transition can only be detected provided that the system satisfies $\eta>0$. Finally, we provide the analytical form of the critical coupling strength:
\begin{flalign}
    J_c(\eta) = \frac{1}{\sqrt{\eta}} \frac{2}{\pi \tilde{g}(0)} = \frac{2}{\pi\sqrt{\eta}} \frac{1}{\frac{1}{D} \sqrt{\frac{2}{\pi}}\sum_{n=0}^{D/2-1}\frac{(2n)!}{2^{2n}(n!)^2}}~, \label{eq:critical_vector}
\end{flalign} which is a function of the symmetry parameter $\eta$. Comparing Eq.~(\ref{eq:critical_vector}) to Eq.~(\ref{eq:critical_pazo}), we observe that both have the same form except for $\tilde{g}(0) \neq g(0)$ due to the higher-dimensional nature of our vector oscillators. Moreover, for $D=2$, we regain the critical coupling strength for the Abelian Kuramoto model in Eq.~(\ref{eq:critical_pazo}).

\begin{figure}[b!]
\centering
\includegraphics[scale=0.25]{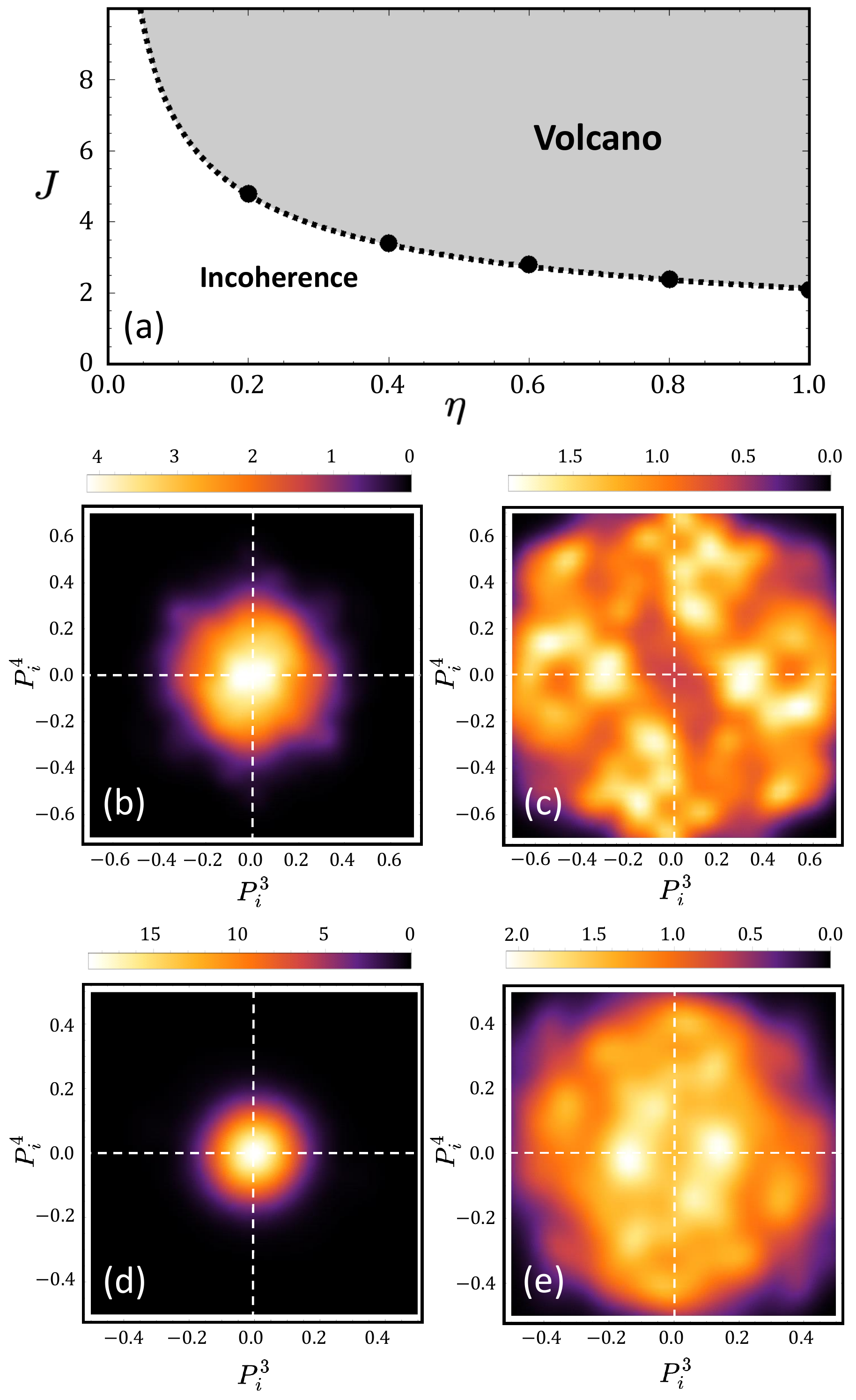}
\caption{(a) Phase diagram obtained from the analytic Eq.~(\ref{eq:critical_vector}). The dashed line indicates the critical coupling strength as a function of the symmetry parameter $\eta$. For comparison, we display critical coupling strengths that are numerically obtained (black solid dots). (b-e) Normalized distributions of components of local fields (\ref{eq:local_field_vector}) for reciprocal interactions with $\eta=1.0$ and (b) $J=1.8$, (c) $J=2.8$ and for nonreciprocal interactions with $\eta=0.8$ and (d) $J=2.0$, (e) $J=2.6$. Note that $J_c=2.12769$ for $\eta=1.0$ and $J_c=2.37883$ for $\eta=0.8$. We perform numerical integration of Eq.~(\ref{eq:vector_governing}) using a fourth-order Runge-Kutta method with a time step of $dt=0.05$ and a system size of $N=500$. After discarding transient steps, we collect the data between $t=1500$ and $t=2000$ for each ensemble, and then compute the average over 100 samples, each having distinct sets of natural frequencies and random interaction matrices.} 
\label{Fig:volcano_4d_overview}
\end{figure}

\section{\label{subsec:4Dmodel}Numerical Results: Volcano Transition for 4D Kuramoto Oscillators}

In Sec.~\ref{sec:theory}, we derived the critical coupling strength (\ref{eq:critical_vector}) as a function of the symmetry parameter $\eta$ in the thermodynamic limit using the higher-dimensional OA ansatz. In this section, we validate our analytical findings through numerical simulations in a system of unit 4-vector oscillators as given in Eq.~(\ref{eq:vector_governing}), considering a large value of $N$ that satisfies $L \ll \log_2 N$ to have a low-rank interaction matrix.

In Fig.~\ref{Fig:volcano_4d_overview} (a), a phase diagram in the $\eta-J$ plane is illustrated together with the analytically determined critical coupling strength $J_c(\eta)$. Akin to the Abelian Kuramoto model~\cite{pazo_volcano}, we observe the divergence of the volcano phase boundary as $\eta \rightarrow 0^+$, a region where no volcano transition is expected to occur. Additionally, the phase diagram shows that a volcano transition can be observed not only with symmetric and reciprocal interactions ($\eta=1$) but also with asymmetric and nonreciprocal interactions ($0<\eta<1$). In Fig.~\ref{Fig:volcano_4d_overview} (b-c), we present normalized distributions for the components of local field vectors, focusing on fully symmetric interactions ($\eta=1$). We numerically assess 2D distributions for the third ($P_i^3$) and fourth ($P_i^4$) components of the local field vectors (\ref{eq:local_field_vector}), both below and above the critical coupling point. Notice that the distributions for the first ($P_i^1$) and second ($P_i^2$) components display the same characteristics (not shown here). As depicted in Fig.~\ref{Fig:volcano_4d_overview} (b), the distribution of local field vectors for $J<J_c$ exhibits a bell-shaped unimodal pattern centered at the origin. This observation confirms that, for $J<J_c$, the most probable value of the local fields is indeed a zero vector. On the contrary, as shown in Fig.~\ref{Fig:volcano_4d_overview} (c), the zero vector is not the most likely local field vector for $J>J_c$; instead, the distribution exhibits its peak at a nonzero vector. This distinctive characteristic, where the behavior differs below and above the critical point, is a hallmark of volcano transitions~\cite{pazo_volcano,strogatz_volcano,Daido_volcano}. Next, we shift our focus to nonreciprocal interactions characterized by $0<\eta<1$. To explore this, we also conducted numerical simulations, however, employing nonreciprocal random interactions ($\eta=0.8$). As illustrated in Fig.~\ref{Fig:volcano_4d_overview} (d-e), similarly, the distribution of local fields follows a bell-shaped unimodal feature for $J<J_c$ whereas it possesses its peaks at nonzero local fields for $J>J_c$, which also indicates a detection of a volcano transition for nonreciprocal interactions.

\begin{figure}[b!]
\centering
\includegraphics[scale=0.25]{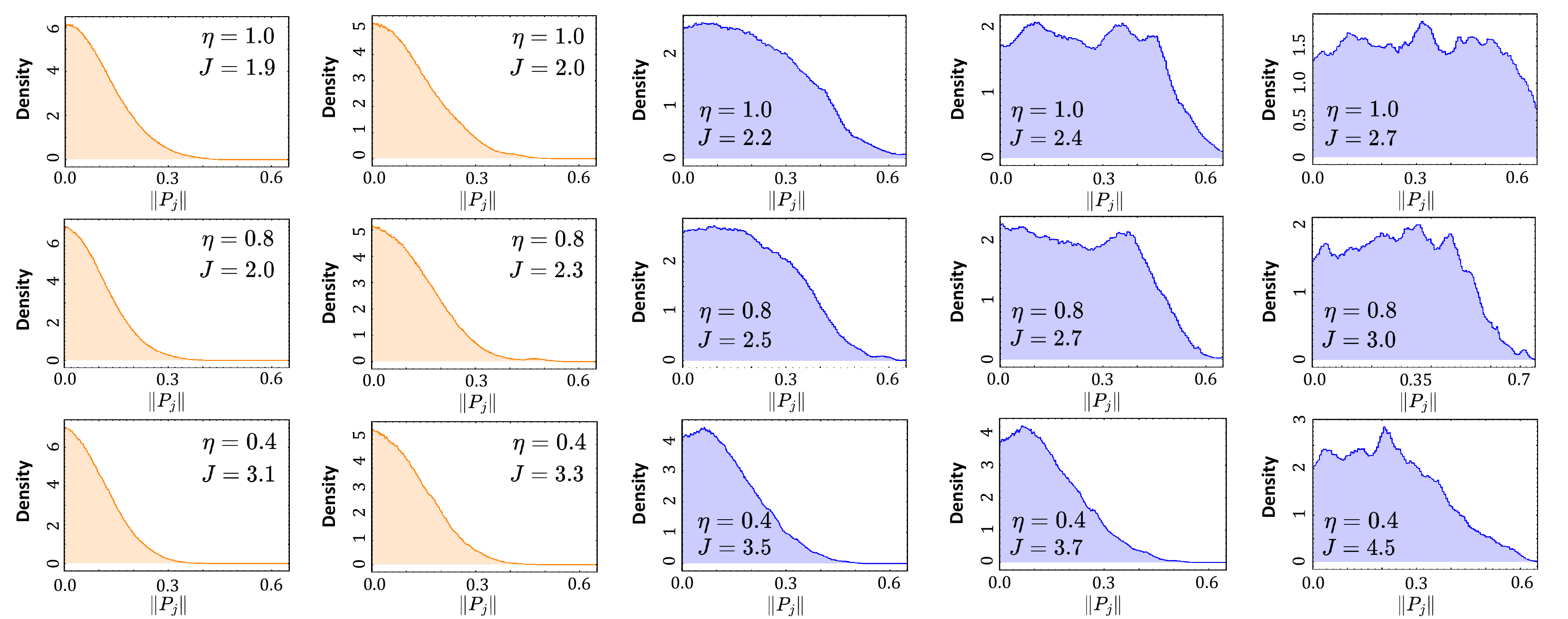}
\caption{Cross-sections of normalized distributions of local fields (\ref{eq:local_field_vector}) for both reciprocal (the first row) and nonreciprocal (the second and third row) random interactions. Note that $J_c=2.12769$ for $\eta=1.0$, $J_c=2.37883$ for $\eta=0.8$, and $J_c=3.36418$ for $\eta=0.4$. Orange-colored distributions are obtained for $J<J_c$ while blue-colored distributions are for $J>J_c$. The numerical methods used are the same as those given in the caption of Fig.~\ref{Fig:volcano_4d_overview}.} 
\label{Fig:volcano_4d_distributions}
\end{figure}

In Fig.~\ref{Fig:volcano_4d_distributions}, cross sections of distributions of local fields are depicted with various coupling strengths for $\eta=1$ (reciprocal), $\eta=0.8$ and $\eta=0.4$ (nonreciprocal). For both reciprocal and nonreciprocal interactions, the distributions are bell-shaped and centered at the origin below the critical coupling strength, which statistically demonstrates that the incoherent state is stable. As the coupling $J$ increases up to $J_c$, this bell-shaped distribution becomes ever broader while the peak at the origin becomes lower. Above the critical point $J>J_c$ the incoherent state is destabilized and the peak of distributions appears at a non-zero point. Finally, we examine the positions of the maximum as a function of the coupling strength in order to illustrate where a volcano transition takes place. 

\begin{figure}[b!]
\centering
\includegraphics[scale=0.5]{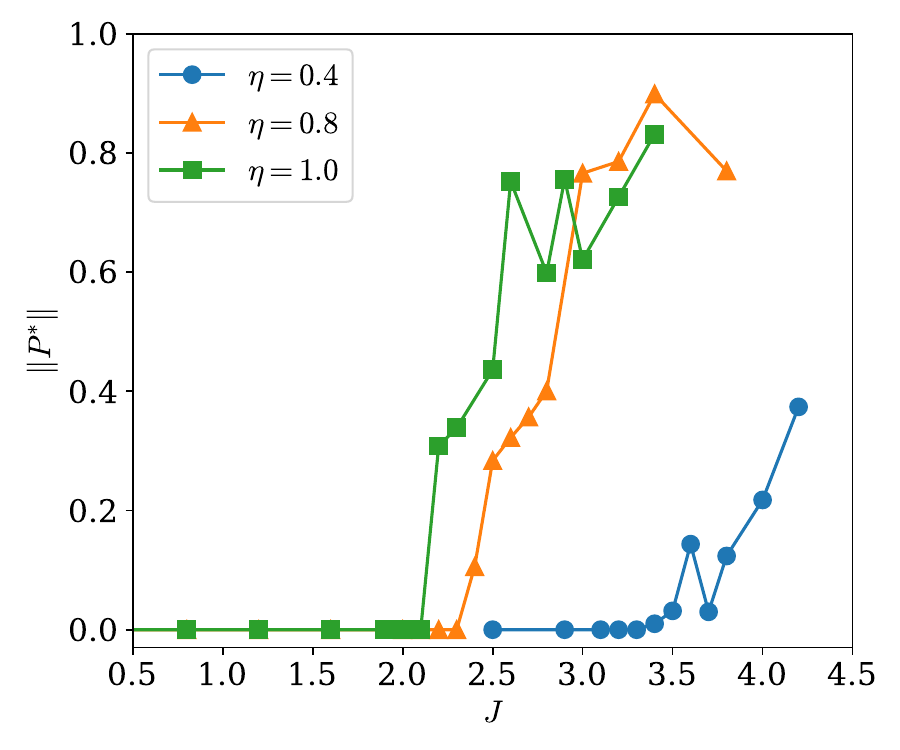}
\caption{Maxima of cross-sections of local field vector distributions as a function of the coupling strength $J$ for different values of $\eta$: Green square ($\eta=1.0$), orange triangle ($\eta=0.8$) and blue circle ($\eta=0.4$). Note that Eq.~(\ref{eq:critical_vector}) predicts $J_c=2.12769$ for $\eta=1.0$, $J_c=2.37883$ for $\eta=0.8$, and $J_c=3.36418$ for $\eta=0.4$. The numerical methods used are the same as those given in the caption of Fig.~\ref{Fig:volcano_4d_overview}.} 
\label{Fig:volcano_4d_peak}
\end{figure}

In Fig.~\ref{Fig:volcano_4d_peak}, the position of the maximum $\| \bold{P}^* \|$ is depicted as a function of the coupling strength $J$ for three different symmetry parameters, within our numerical capacity. The maximum $\| \bold{P}^*\|$ stays near zero for $J<J_c$ where the incoherent state is stable. However, it bifurcates off the zero value close to the value of $J_c$ analytically determined in Eq.~(\ref{eq:critical_vector}). Note that beyond the critical point, we observe certain fluctuations as the coupling strength increases. This observation might arise from either the limited number of ensembles or the limited number of oscillators.

\begin{figure}[t!]
\centering
\includegraphics[scale=0.35]{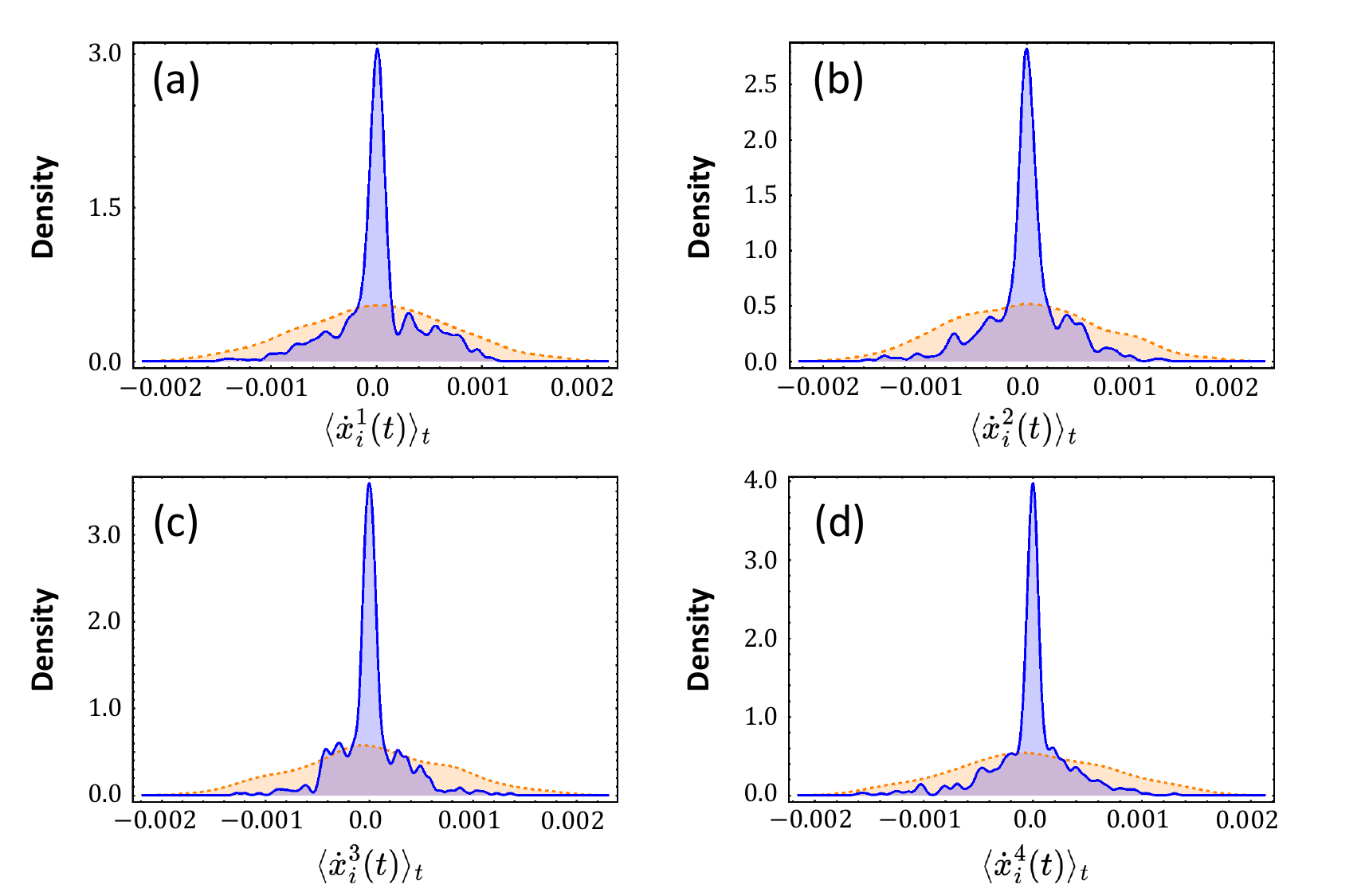}
\caption{Distribution of time-averaged velocities of each component of vector oscillators: (a) $\langle \dot{x}_i^1 \rangle_t$, (b) $\langle \dot{x}_i^2 \rangle_t$, (c) $\langle \dot{x}_i^3 \rangle_t$ and (d) $\langle \dot{x}_i^4 \rangle_t$. A blue histogram: $J=2.8 > J_c$; an orange histogram: $J=1.8 <J_c$. All figures are obtained from reciprocal interactions ($\eta=1$). However, similar results for nonreciprocal interactions with $\eta<1$ (results not shown here) can be found. We perform numerical integration of Eq.~(\ref{eq:vector_governing}) using a fourth-order Runge-Kutta method with a time step of $dt=0.05$ and a system size of $N=500$. We average the instantaneous velocities of each component for $\Delta t=1000$ up to $T_\text{max}=3000$.} 
\label{Fig:volcano_4d_vel}
\end{figure}

As discussed in Sec.~\ref{sec:theory}, the volcano transition is characterized by the loss of stability of the incoherent and the emergence of partial synchrony. To illustrate this phenomenon, we examine the time-averaged velocity of each component of the oscillators, defined as
$    \langle \dot{x}_i^a(t) \rangle_t = \frac{1}{T} \int_0^T \dot{x}_{i}^a(t') dt' $
for $a=1,2,3,4$. In Fig.~\ref{Fig:volcano_4d_vel}, we present density distributions of time-averaged velocities for each component, both below and above the transition point $J_c$. Below the transition (orange), incoherent states are evident, showing no substantial locking behavior and exhibiting a relatively broad distribution around zero. Conversely, above the volcano transition (blue), a significant portion of the oscillators is locked, resulting in a density distribution of time-averaged velocities that is highly concentrated around zero. This observation corroborates the fact that the incoherent state indeed becomes destabilized, giving rise to a partially locked state for $J>J_c$. It is noteworthy that a similar outcome was observed in a system of Abelian Kuramoto oscillators in Ref.~\cite{strogatz_volcano}.

For a more detailed analysis of the properties of observable dynamical states, we explore the degree of correlation between each oscillator $\bold{x}_i$ and the local field vector $\bold{P}_j$. As previously noted, the quenched random interaction matrix $\bold{M}$ encompasses both attractive ($M_{ij}>0$) and repulsive ($M_{ij}<0$) couplings between oscillators. Consequently, in a partially locked state for $J>J_c$, an oscillator $\bold{x}_i$ tends to align with the local field vector $\bold{P}_j$ when it is coupled attractively with oscillator $\bold{x}_j$, i.e., $M_{ij}>0$. Conversely, in the case of repulsive coupling, the oscillator $\bold{x}_i$ tends to be in anti-alignment with the local field vector $\bold{P}_j$~\cite{strogatz_volcano}. In this analysis, we explore the cosine similarity between an oscillator $\bold{x}_i$ and the local field vector $\bold{P}_j$ for all $i,j \in [N]$ defined by
\begin{equation}
    S_c(\bold{x}_i,\bold{P}_j) := \frac{\langle \bold{x}_i, \bold{P}_j \rangle}{ \| \bold{x}_i \| \|\bold{P}_j \|} ~.\label{eq:angle}
\end{equation}

\begin{figure}[t!]
\centering
\includegraphics[scale=0.25]{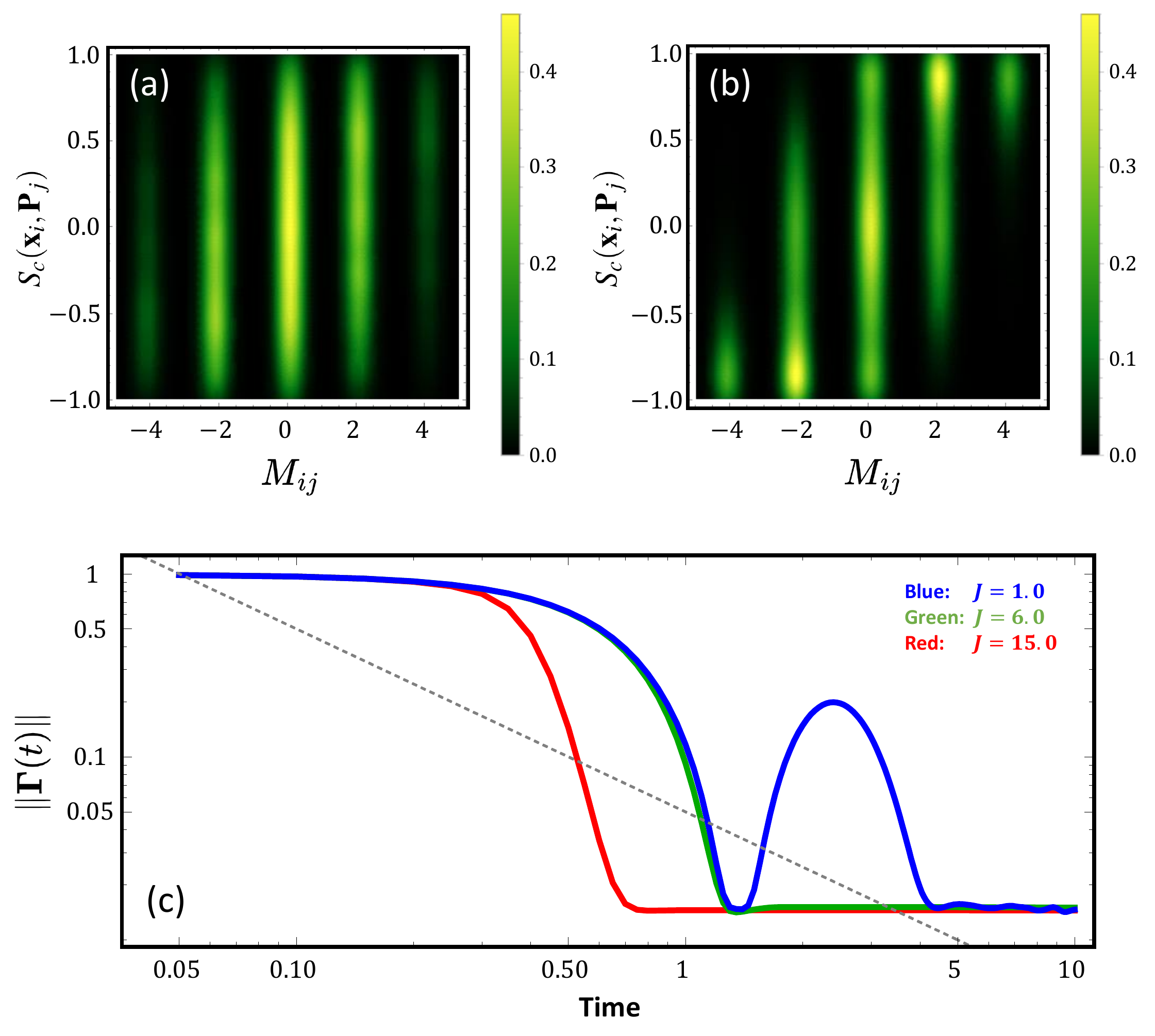}
\caption{Densities of $S_c(\bold{x}_i,\bold{P}_j)$ (see Eq.~(\ref{eq:angle})) are plotted against the interaction matrix element $M_{ij}$ for (a) $J=1.8<J_c$ and (b) $J=2.8>J_c$. All figures are obtained for reciprocal interactions ($\eta=1$). However, similar results for nonreciprocal interactions with $\eta<1$ (results not shown here) can be found. We perform numerical integration of Eq.~(\ref{eq:vector_governing}) using a fourth-order Runge-Kutta method with a time step of $dt=0.05$ and a system size of $N=4000$. After discarding transient steps, we collect the data between $t=1000$ and $t=2000$. (c) Time evolution of the norm of the global Kuramoto order parameter (\ref{eq:global_order}) for $\eta=1.0$ in a log-log scale. Gray dashed guideline indicates an algebraic relaxation. All the considered cases exhibit an exponential relaxation: $J=1.0 <J_c$ (blue), $J=6.0 >J_c$ (green), and $J=15.0 >J_c$ (red). }  
\label{Fig:volcano_4d_cor}
\end{figure}

Note that when oscillator $i$ is aligned with the local field $j$, then $S_c(\bold{x}_i,\bold{P}_j) = 1$, indicating that the angle between them is 0. In contrast, $S_c(\bold{x}_i,\bold{P}_j) = -1$ implies that the angle between them is $\pm \pi$, i.e., they are in anti-alignment. In Fig.~\ref{Fig:volcano_4d_cor}, we display the densities of $S_c(\bold{x}_i,\bold{P}_j)$ as a function of the interaction matrix element $M_{ij}$ for both (a) $J<J_c$ and (b) $J>J_c$. Below the volcano transition, we observe that the local field vectors are not correlated to the oscillators that characterize the incoherent state. This can be noticed from the distribution of alignments, which shows vertical stripes~\cite{strogatz_volcano}, as depicted in Fig.~\ref{Fig:volcano_4d_cor} (a). However, as shown in Fig.~\ref{Fig:volcano_4d_cor} (b), above the critical point, the density of $S_c(\bold{x}_i,\bold{P}_j)$ versus $M_{ij}$ exhibits peaks at approximately $S_c(\bold{x}_i,\bold{P}_j)=1$ for the attractive coupling $M_{ij}>0$ and at around $S_c(\bold{x}_i,\bold{P}_j)=-1$ for the repulsive coupling $M_{ij}<0$. This observation demonstrates how the oscillators are aligned with respect to the local fields in the emergence of a partially locked state (a volcano phase)~\cite{strogatz_volcano}.

In previous studies~\cite{strogatz_volcano,glassy1,glassy2,glassy3,glassy4}, it was conjectured that an oscillator glass might be characterized by nonexponential or algebraic relaxation of the global Kuramoto order parameter $\Gamma(t)$ defined in Eq.~(\ref{eq:global_order_pazo}). To investigate the decay of the global Kuramoto order parameter in our system, we tracked $\| \bm{\Gamma}(t)\|$ as given in Eq.~(\ref{eq:global_order}) over time, commencing from an initially synchronized state. In Fig.~\ref{Fig:volcano_4d_cor} (c), the decay of the norm of the global Kuramoto order parameter is shown as a function of time with different coupling strengths. For both cases of $J<J_c$ and $J>J_c$, we observed only an exponential relaxation of the norm of the center of mass vector as time passes. This outcome is in line with the findings in the Abelian Kuramoto model. In Ref.~\cite{strogatz_volcano}, the authors showed that a volcano phase exhibits an exponential relaxation of the global order parameter for a low-rank interaction matrix, i.e., $L \ll \log_2N$. Here, we arrive at the same conclusion: The presence of volcano phases in a system of non-Abelian Kuramoto oscillators (\ref{eq:vector_governing}) does not necessarily indicate the existence of a glassy state for $L \ll \log_2N$.

\section{\label{subsec:3Dmodel}Absence of Volcano Transition for 3D Kuramoto Oscillators}

So far, we have explored a volcano transition that occurs in a system of heterogeneous unit 4-vector oscillators with reciprocal and nonreciprocal random interactions. Various studies have highlighted distinctions between even and odd dimensions in the collective dynamics of systems comprising higher-dimensional oscillators. For example, globally coupled heterogeneous oscillators show different characteristics in terms of the onset of partial synchronization. For odd dimensions, a partially locked state discontinuously occurs, no matter how small the global coupling strength is~\cite{PRX_generalized,barioni2,g_KM7}. Furthermore, when we consider odd dimensions, higher-dimensional, identical oscillators with a phase-lag rotation matrix are known to possess additional complete synchronization as fixed points at the north and south poles, regardless of the phase-lag parameter~\cite{Lee_2023}. This particular distinction arises due to a specific direction of a rotation in an odd dimension. An even-dimensional rotation is an isoclinic rotation consisting of planes of rotations. On the contrary, for odd dimensions, there is a so-called rotational axis perpendicular to the planes of rotation. In many cases, oscillators are inclined to this axis such that one can have complete synchronization at the poles~\cite{Lee_2023}, or a partially locked state along the natural frequency vector~\cite{PRX_generalized,Lohe_2009,barioni2,g_KM7}, regardless of a given parameter.

In previous studies on volcano transitions~\cite{strogatz_volcano,pazo_volcano,Daido_volcano,daido_volcano2}, a system of Abelian Kuramoto oscillators was investigated with random frustrated interactions. In the context of the current paper, such Abelian Kuramoto oscillators are 2-dimensional vector oscillators on the unit circle. Thus, one can expect a volcano transition from an incoherent state to a volcano phase, as for 4D oscillators in the previous sections. In this section, we demonstrate that no volcano transition occurs for odd dimensions since oscillators are partially locked along the average natural frequency vector, irrespective of the global parameter $J>0$. We study a system of unit 3-vectors with the same random interactions. Note that equation (\ref{eq:Lohe_governing}) for $\textrm{SU}(2)$ corresponds to the unit 4-vector model on $S^3$. To consider a system of unit 3-vector oscillators on $S^2$, Lohe restricted the system to follow
\begin{equation}
    \textrm{i} \dot{U}_i {U_i}^\dagger = H_i - U_i H_i {U_i}^\dagger - \frac{\textrm{i}J}{2N}\sum_{j=1}^N M_{ij} \bigg( U_i {U_j}^\dagger - U_j {U_i}^\dagger \bigg) ,\label{eq:Lohe_governing_3d}
\end{equation} where $U_i \in \textrm{SU}(2)$ for $i \in [N]$. In this case, the chiral covariance is given by $S=T$~\cite{Lohe_2009}. Then, oscillators in Eq.~(\ref{eq:Lohe_governing_3d}) can be parameterized by the same method as given in Eq.~(\ref{eq:parameterization_unitary}), however, setting $x_i^4\equiv 0$ identically. Thus, one obtains the following system of unit 3-vectors~\cite{barioni1,Lohe_2009} for all $i \in [N]$:
\begin{equation}
    \dot{\bold{x}}_i = \bm{\omega}_i \times \bold{x}_i + \frac{J}{N} \sum_{j=1}^{N}M_{ij} \big( \bold{x}_j - \langle \bold{x}_i, \bold{x}_j \rangle \bold{x}_i \big) ~, \label{eq:3d_governing}
\end{equation}  
where the natural frequency vector reads $\bm{\omega}_i =(\omega_i^1,\omega_i^2,\omega_i^3)$, each component of which is selected randomly and independently from an unimodal distribution~\cite{Lohe_2009}. The OA ansatz in Sec.~\ref{subsec:higher_OA} can be applied to this system with $D=3$, which leads to the following evolution equations of the OA vector and its norm~\cite{barioni1}:
\begin{flalign}
    &\frac{d}{dt} \boldgreek{\bm{\rho}} = \bm{\omega} \times \boldgreek{\bm{\rho}} + \frac{J}{2}\big( 1+ \| \boldgreek{\bm{\rho}} \|^2 \big) \bold{P} - J \langle \boldgreek{\bm{\rho}}, \bold{P} \rangle \boldgreek{\bm{\rho}} \notag \\
     &\frac{d}{dt} \| \boldgreek{\bm{\rho}} \| = \frac{J}{2}(1-\|\boldgreek{\bm{\rho}}\|^2) \langle \hat{\boldgreek{\bm{\rho}}}, \bold{P} \rangle . \label{eq:OA_3d_vector}
\end{flalign} We now look for an equilibrium solution of Eq.~(\ref{eq:OA_3d_vector}). The second equation indicates that an equilibrium solution should be either $\|\boldgreek{\bm{\rho}}\|=1$ or $ \hat{\boldgreek{\bm{\rho}}} \perp \bold{P}$. When we consider the coupling strength $J$ as a very small parameter, then $\frac{d}{dt} \boldgreek{\bm{\rho}}  \sim \bm{\omega} \times \boldgreek{\bm{\rho}} $ as $J \rightarrow 0^+$, which thus implies that an equilibrium should satisfy $\boldgreek{\bm{\rho}} \sim \pm \bm{\omega}$. Employing small perturbations $\boldgreek{\bm{\rho}} = \pm \bm{\omega} + \delta\boldgreek{\bm{\rho}} $ and $\| \boldgreek{\bm{\rho}} \| = 1+\delta\rho$, one can show that the solution $\boldgreek{\bm{\rho}} \sim \pm \bm{\omega}$ is stable for $\langle \hat{\boldgreek{\bm{\rho}}}, \bold{P} \rangle>0$ since the perturbation is governed by $\frac{d}{dt}\delta\rho = -\frac{J}{2}\langle \hat{\boldgreek{\bm{\rho}}}, \bold{P} \rangle  \delta\rho$~\cite{barioni1}. Thus, we draw a conclusion that there is no volcano transition for $D=3$: The locked state, i.e., either $\boldgreek{\bm{\rho}} \sim \bm{\omega}$ or $\boldgreek{\bm{\rho}} \sim -\bm{\omega}$, becomes already stable for $J \rightarrow 0^+$, regardless of the value of $\eta$. 

\begin{figure}[t!]
\centering
\includegraphics[scale=0.37]{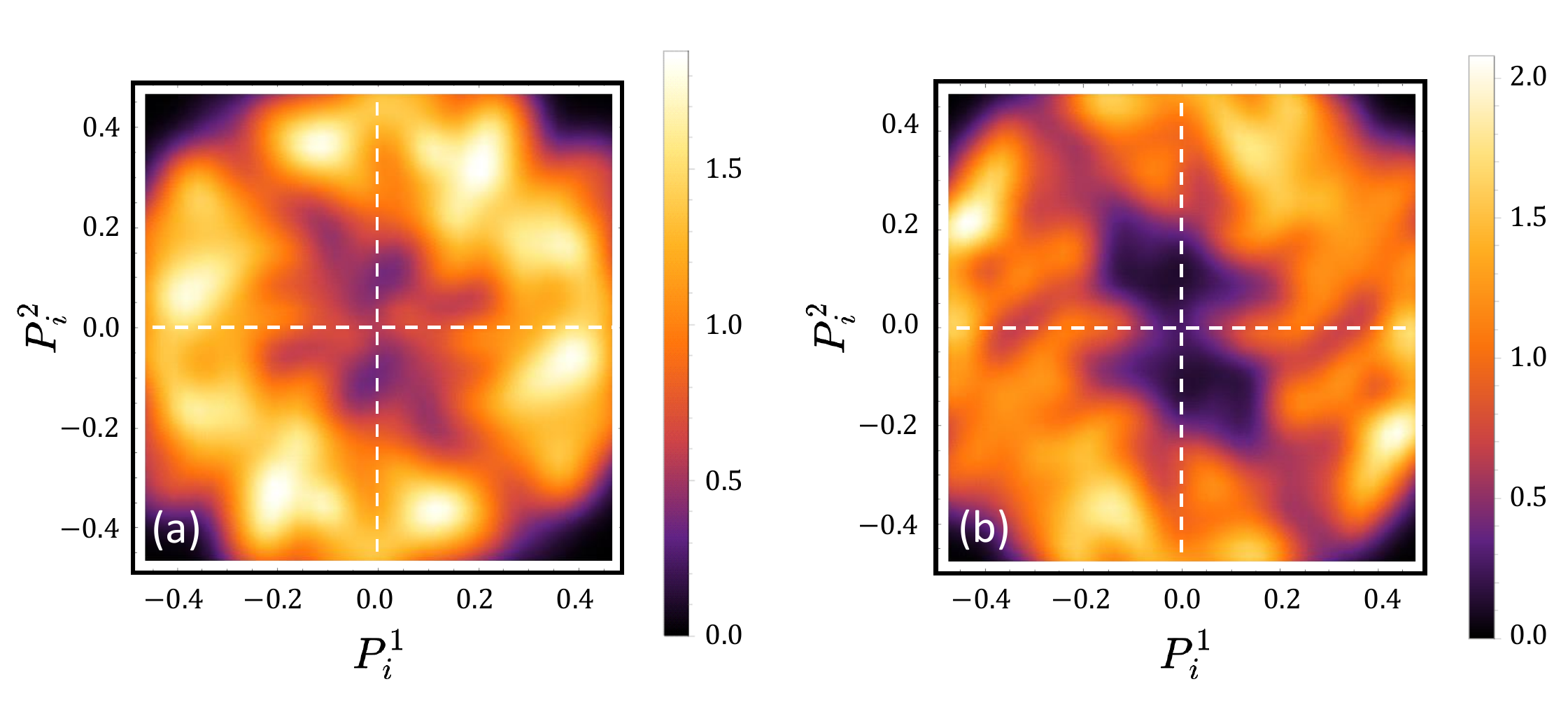}
\caption{Distributions of components of the local fields (\ref{eq:local_field_vector}) for (a) $J=0.01$ and (b) $J=0.05$. The numerical methods used are the same as those given in the caption of Fig.~\ref{Fig:volcano_4d_overview}.} 
\label{Fig:3d_volcano}
\end{figure}

This conjecture is verified by numerically solving Eq.~(\ref{eq:3d_governing}) for large $N$. In Fig.~\ref{Fig:3d_volcano}, distributions of components of local fields are depicted for very small coupling strengths: (a) $J=0.01$ and (b) $J=0.05$. For both cases, we observe a volcano-shaped distribution of local fields. The most probable value of the local fields does not occur at the origin but rather the peak of the distribution is detected at a non-zero vector. This numerical observation is consistent with the above analytical conjecture that no volcano transition can be observed in a system of unit 3-vector oscillators on the 2-sphere.

\section{\label{sec:Conlcusion}Summary and Outlook}

In this paper, we generalized a system of Kuramoto phase oscillators to higher-dimensional Kuramoto vector oscillators with both reciprocal and nonreciprocal frustrated interactions proposed in Ref.~\cite{pazo_volcano}. This generalization also predicted a volcano transition in which the incoherent state becomes unstable and a partially locked state emerges. We derived an analytical expression of the critical coupling strength at which the volcano transition occurs using the so-called higher-dimensional OA ansatz in the thermodynamic limit~\cite{barioni2}. With numerical simulations of large-sized ensembles of oscillators, we validated our analytical results for 4D oscillators. The obtained collective behaviors exhibited distinct dynamical characteristics below and above the volcano transition, respectively. A significant portion of oscillators above the volcano transition becomes entrained, and their vectors align or anti-align with local field vectors, depending on whether the interactions between oscillators are attractive or repulsive. In contrast, the incoherent states below the threshold point do not exhibit these characteristics. Furthermore, we proposed the absence of a volcano transition in a system of 3D Kuramoto oscillators. This conjecture was supported by our observation of a stable partially locked state as the coupling strength approached zero from the positive side. We also confirmed this conjecture through numerical simulations. Note that another study on higher-dimensional phase oscillators with random frustrated interactions was recently reported, where the authors considered an equilibrium in the mean-field equation for vector spin models on random networks with high connectivity, featuring an arbitrary degree distribution and links with random weights~\cite{metz2022mean}.

In fact, our analytical outcome (\ref{eq:critical_vector}) is not confined to 4D Kuramoto oscillators alone. Therefore, one has the flexibility to numerically substantiate this conclusion using an ensemble of oscillators with any even dimensionality, such as $D=6,8, \cdots$. This could provide further insights into how the dimensionality of the oscillators influences observable collective dynamics. Similarly, the conjecture of the absence of a volcano transition can be extended to encompass investigations for any odd-dimensional oscillators in future research. Moreover, in this paper, we focused solely on low-rank interaction matrices for the sake of analytical convenience. Nevertheless, just as in Refs.~\cite{strogatz_volcano,pazo_volcano,Daido_volcano}, one can employ full-ranked interaction matrices to investigate a volcano transition and its relation with glassy oscillator states, which still remains controversial. 

Recently, generalized Kuramoto oscillators have received considerable attention, with a specific emphasis on the dynamical and spectral properties of their collective behaviors. On the one hand, there has been a trend toward exploring higher-dimensional generalizations to investigate phenomena like synchronization or chimera states~\cite{Tanaka_2014,g_KM1,g_KM2,g_KM3,g_KM4,g_KM5,g_KM6,g_KM7,short_chimera,Lee_2023}. On the other hand, there has been a growing interest in the complexification of oscillator dynamics~\cite{complexified1,complexified2,Seung-Yeal,quaternion}. While these two generalizations possess distinct characteristics, it is worth considering future investigations into any generalization of Kuramoto oscillators to explore collective dynamics within a variety of interaction setups. This could encompass scenarios such as systems with other random frustrated interactions than those discussed here, oscillators arranged on a ring geometry, or within graph topologies, as these could offer avenues for a closer approximation to real-world systems.

\section*{Data Availability Statement}
The data that support the findings of this study are available from the corresponding author upon reasonable request.

\section*{Acknowledgments}
S.L. and K.K. have been supported by the Deutsche Forschungsgemeinschaft (DFG project KR1189/18-2). Y.J. and S. -W.S. have been supported by the National Research Foundation (NRF) of Korea through the Grant Numbers. NRF-2023R1A2C1007523 (S.-W.S.).

\section*{Author Contributions}
\textbf{Seungjae Lee:} Conceptualization (lead); Formal analysis (lead); Investigation (lead); Writing - original draft (lead); Software (lead); Writing - review \text{\&} editing (equal); Visualization (equal). \textbf{Yeonsu Jeong:} Investigation (support); Software (support); Visualization (equal); Writing - review \text{\&} editing (equal). \textbf{Seung-Woo Son:} Supervision (equal); Funding acquisition (lead); Writing - review \text{\&} editing (equal).  \textbf{Katharina Krischer:} Supervision (equal); Funding acquisition (lead); Writing - review \text{\&} editing (equal).

\vskip 1cm 
\section*{References}
\bibliographystyle{unsrt}
\bibliography{ref.bib}

\end{document}